\documentclass[sigconf,screen]{acmart}

\renewcommand\footnotetextcopyrightpermission[1]{}
\copyrightyear{2024} 
\acmYear{2024} 
\setcopyright{rightsretained} 
\acmConference[ICSE-NIER'24]{New Ideas and Emerging Results }{April 14--20, 2024}{Lisbon, Portugal}
\acmBooktitle{New Ideas and Emerging Results (ICSE-NIER'24), April 14--20, 2024, Lisbon, Portugal}\acmDOI{10.1145/3639476.3639769}
\acmISBN{979-8-4007-0500-7/24/04}

\usepackage{subcaption}
\usepackage{multirow}
\usepackage{colortbl}
\usepackage{svg}
\usepackage{microtype}
\usepackage{url}
\usepackage{tcolorbox}

\usepackage{macros}
\newlength{\textfloatsepsave}
\usepackage{listings}
\usepackage{xcolor}

\colorlet{punctcolor}{red!60!black}
\colorlet{desccolor}{green!60!black}
\colorlet{altcolor}{blue!60!black}
\colorlet{kwcolor}{teal!60!black}
\definecolor{keywordcolor}{rgb}{0.13, 0.29, 0.53}



\author{Myeongsoo Kim}
\affiliation{
  \institution{Georgia Institute of Technology}
  \city{Atlanta}
  \state{Georgia}
  \country{USA}
}
\email{mkim754@gatech.edu}

\author{Tyler Stennett}
\affiliation{
  \institution{Georgia Institute of Technology}
  \city{Atlanta}
  \state{Georgia}
  \country{USA}
}
\email{tstennett3@gatech.edu}

\author{Dhruv Shah}
\affiliation{
  \institution{Georgia Institute of Technology}
  \city{Atlanta}
  \state{Georgia}
  \country{USA}
}
\email{dshah374@gatech.edu}

\author{Saurabh Sinha}
\affiliation{
  \institution{IBM Research}
  \city{Yorktown Heights}
  \state{New York}
  \country{USA}
}
\email{sinhas@us.ibm.com}

\author{Alessandro Orso}
\affiliation{
  \institution{Georgia Institute of Technology}
  \city{Atlanta}
  \state{Georgia}
  \country{USA}
}
\email{orso@cc.gatech.edu}

\begin{document}

\title{Leveraging Large Language Models to Improve REST API Testing}

\begin{abstract}
The widespread adoption of REST APIs, coupled with their growing complexity and size, has led to the need for automated REST API testing tools. Current tools focus on the structured data in REST API specifications but often neglect valuable insights available in unstructured natural-language descriptions in the specifications, which leads to suboptimal test coverage. Recently, to address this gap, researchers have developed techniques that extract rules from these human-readable descriptions and query knowledge bases to derive meaningful input values. However, these techniques are limited in the types of rules they can extract and prone to produce inaccurate results. This paper presents RESTGPT, an innovative approach that leverages the power and intrinsic context-awareness of Large Language Models (LLMs) to improve REST API testing. RESTGPT takes as input an API specification, extracts machine-interpretable rules, and generates example parameter values from natural-language descriptions in the specification. It then augments the original specification with these rules and values. Our evaluations indicate that RESTGPT outperforms existing techniques in both rule extraction and value generation. Given these promising results, we outline future research directions for advancing REST API testing through LLMs. 
\end{abstract}

\maketitle
\pagestyle{plain}

\section{Introduction}
\label{sec:introduction}
In today's digital era, web applications and cloud-based systems have become ubiquitous, making REpresentational State Transfer (REST) Application Programming Interfaces (APIs) pivotal elements in software development~\cite{richardson2013restful}. REST APIs enable disparate systems to communicate and exchange data seamlessly, facilitating the integration of a wide range of services and functionalities~\cite{fielding2000architectural}. As their intricacy and prevalence grow, effective testing of REST APIs has emerged as a significant challenge~\cite{kim2022automated,zhang2022open,golmohammadi2023survey}.

Automated REST API testing tools (e.g.,~\cite{arcuri2019restful, Corradini2022, atlidakis2019restler, karlsson2020quickrest, martin2021restest, karlsson2020automatic, zac2022schemathesis, wu2022combinatorial, kim2023reinforcement, dredd, tcases}) primarily derive test cases from API specifications~\cite{openapi, swagger, raml, apiblueprint}. Their struggle to achieve high code coverage~\cite{kim2022automated} often stems from difficulties in comprehending the semantics and constraints present in parameter names and descriptions~\cite{kim2022automated,kim2023enhancing,alonso2022arte}. To address these issues, assistant tools have been developed. These tools leverage Natural Language Processing (NLP) to extract constraints from parameter descriptions~\cite{kim2023enhancing} and query parameter names against databases~\cite{alonso2022arte}, such as DBPedia~\cite{bizer2009dbpedia}. However, attaining high accuracy remains a significant challenge for these tools. Moreover, they are limited in the types and complexity of rules they can extract. 

This paper introduces RESTGPT, a new approach that harnesses Large Language Models (LLMs) to enhance REST API specifications by identifying constraints and generating relevant parameter values. Given an OpenAPI Specification~\cite{openapi}, 
RESTGPT augments it by deriving constraints and example values.
Existing approaches such as NLP2REST~\cite{kim2023enhancing} require a validation process to improve precision, which involves not just the extraction of constraints but also executing requests against the APIs to dynamically check these constraints. Such a process demands significant engineering effort and a deployed service instance, making it cumbersome and time-consuming.
In contrast, RESTGPT achieves higher accuracy without requiring expensive validation. Furthermore, unlike ARTE~\cite{alonso2022arte}, RESTGPT excels in understanding the context of a parameter name based on an analysis of the parameter description, thus generating more contextually relevant values.


Our preliminary results demonstrate the significant advantage of our approach over existing tools. Compared to NLP2REST without the validation module, our method improves precision from 50\% to 97\%. Even when compared to NLP2REST equipped with its validation module, our approach still increases precision from 79\% to 97\%. Additionally, RESTGPT successfully generates both syntactically and semantically valid inputs for 73\% of the parameters over the analyzed services and their operations, a considerable improvement over ARTE, which could generate valid inputs for 17\% of the parameters only. Given these encouraging results, we outline a number of research directions for leveraging LLMs in other ways for further enhancing REST API testing.


\begin{figure}[t]
  \centering
  \begin{minipage}{\linewidth}
    \begin{lstlisting}[
        language=json,
        % basicstyle=\tiny,
        belowskip=-0.7\baselineskip
        ]
/institutions:
  @\color{keywordcolor}get@:
    @\color{keywordcolor}operationId@: searchInstitutions
    @\color{keywordcolor}produces@:
      - application/json
    @\color{keywordcolor}parameters@:
      - @\color{keywordcolor}name@: filters
        @\color{keywordcolor}in@: query
        @\color{keywordcolor}required@: false
        @\color{keywordcolor}type@: string
        @\color{keywordcolor}description@: The filter for the bank search.
            Examples:
            * Filter by State name 
            `STNAME:\"West Virginia\"`
            * Filter for any one of multiple State names 
            `STNAME:(\"West Virginia\",\"Delaware\")`
      - @\color{keywordcolor}name@: sort_order
        @\color{keywordcolor}in@: query
        @\color{keywordcolor}required@: false
        @\color{keywordcolor}type@: string
        @\color{keywordcolor}description@: Indicator if ascending (ASC) or descending (DESC)
    @\color{keywordcolor}responses@:
      @\color{keywordcolor}'200'@:
        @\color{keywordcolor}description@: successful operation
        @\color{keywordcolor}schema@:
          @\color{keywordcolor}type@: object
    \end{lstlisting}
  \end{minipage}
  \caption{A part of FDIC Bank Data's OpenAPI specification.}
  \label{fig:oas_example}
\end{figure}
\section{Background and Motivating Example}
\label{background}


\subsection{REST APIs and OpenAPI Specification}

REST APIs are interfaces built on the principles of Representational State Transfer (REST), a design paradigm for networked applications~\cite{fielding2000architectural}. Designed for the web, REST APIs facilitate data exchange between clients and servers through predefined endpoints primarily using the HTTP protocol~\cite{rodriguez2008restful, tilkov2007brief}. Each client interaction can include headers and a payload, while the corresponding response typically contains headers, content, and an HTTP status code indicating the outcome.

OpenAPI Specification (OAS)~\cite{openapi} is arguably the industry standard for defining RESTful API interfaces. It offers the advantage of machine-readability, supporting automation processes, while also presenting information in a clear, human-readable format. Key features of OAS include the definition of endpoints, the associated HTTP methods, expected input parameters, and potential responses. As an example, Figure~\ref{fig:oas_example} shows a portion of the FDIC Bank Data's API specification. This part of the specification illustrates how one might query information about institutions. It also details an expected response, such as the 200 status code, which indicates a successfully processed scenario.


\subsection{REST API Testing and Assistant Tools}

Automated REST API testing tools~\cite{arcuri2019restful, Corradini2022, atlidakis2019restler, karlsson2020quickrest, martin2021restest, karlsson2020automatic, zac2022schemathesis, wu2022combinatorial, kim2023reinforcement, tcases} derive test cases from widely-accepted specifications, primarily OpenAPI~\cite{openapi}. However, these tools often struggle to achieve comprehensive coverage~\cite{kim2022automated}. A significant reason for this is their inability to interpret human-readable parts of the specification~\cite{kim2022automated,kim2023enhancing}. For parameters such as \texttt{\small filters} and \texttt{\small sort\_order} shown in Figure~\ref{fig:oas_example}, testing tools tend to generate random string values, which are often not valid inputs for such parameters.

In response to these challenges, assistant tools have been introduced to enhance the capabilities of these testing tools. For instance, ARTE~\cite{alonso2022arte} taps into DBPedia~\cite{bizer2009dbpedia} to generate relevant parameter example values. Similarly, NLP2REST applies natural language processing to extract example values and constraints from descriptive text portions of the specifications~\cite{kim2023enhancing}.


\subsection{Large Language Model}

Large Language Models (LLMs)~\cite{openai2023gpt4,touvron2023llama,google2023bard} represent a transformative leap in the domains of natural language processing (NLP) and Machine Learning. Characterized by their massive size, often containing billions of parameters, these models are trained on vast text corpora to generate, understand, and manipulate human-like text~\cite{radford2019better}. The architecture behind LLMs are primarily transformer-based designs~\cite{vaswani2017attention}. Notable models based on this architecture include GPT (Generative Pre-trained Transformer)~\cite{radford2018improving}, designed mainly for text generation, and BERT (Bidirectional Encoder Representations from Transformers)~\cite{devlin2019bert}, which excels in understanding context. These models capture intricate linguistic nuances and semantic contexts, making them adept at a wide range of tasks from text generation to answering questions.

\begin{figure}[t]
\centering
\includegraphics[width=\columnwidth]{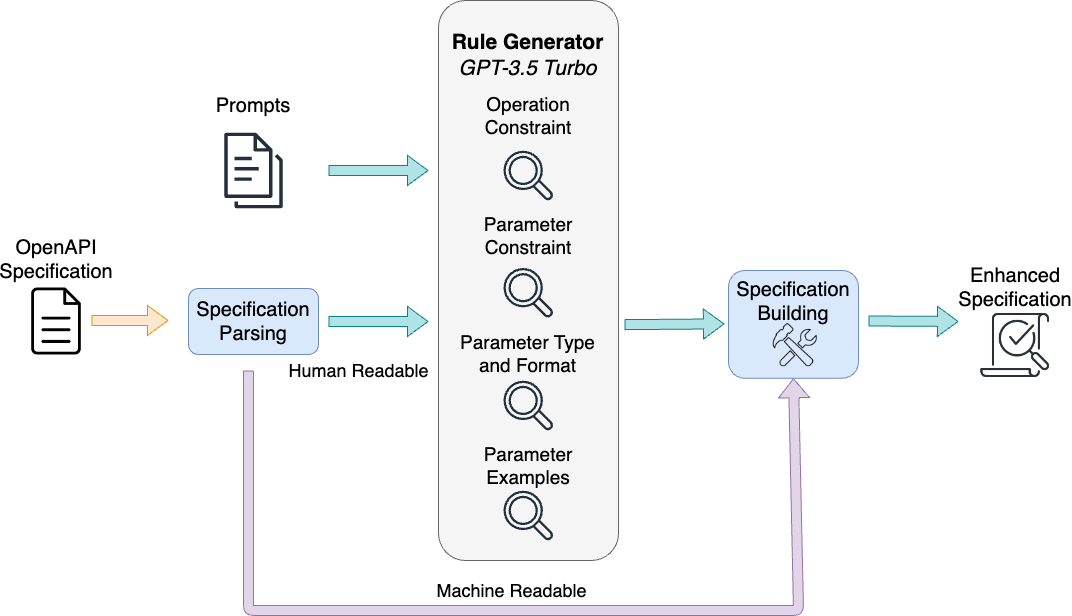}
\caption{Overview of our approach.}
\label{fig:overview}
\end{figure}

\subsection{Motivating Example}

The OpenAPI specification for the Federal Deposit Insurance Corporation (FDIC) Bank Data's API, shown in Figure~\ref{fig:oas_example}, serves to offer insights into banking data. Using this example, we highlight the challenges in parameter value generation faced by current REST API testing assistant tools and illustrate how RESTGPT addresses these challenges.


\begin{enumerate}
    \item \textbf{Parameter \texttt{\small filters}}: Although the description provides guidance on how the parameter should be used, ARTE's dependency on DBPedia results in no relevant value generation for \texttt{\small filters}. NLP2REST, with its keyword-driven extraction, identifies examples from the description, notably aided by the term ``example''. Consequently, patterns such as  \texttt{\small STNAME: "West Virginia"} and \texttt{\small STNAME:\,("West Virginia", "Delaware")} are accurately captured.

    \item \textbf{Parameter \texttt{\small sort\_order}}: Here, both tools exhibit limitations. ARTE, while querying DBPedia, fetches unrelated values such as ``List of colonial heads of Portuguese Timor'', highlighting its contextual inadequacy. In the absence of identifiable keywords, NLP2REST fails to identify ``ASC'' or ``DESC'' as potential values.
\end{enumerate}

In contrast to these tools, RESTGPT is much more effective: with a deeper semantic understanding, RESTGPT accurately discerned that the \texttt{\small filters} parameter was contextualized around state names tied to bank records, and generated test values such as \texttt{\small STNAME: "California"} and multi-state filters such as \texttt{\small STNAME: ("California", "New York")}. Also, it successfully identifies the values ``ASC'' or ``DESC'' from the description of the \texttt{\small sort\_order} parameter. This example illustrates RESTGPT's superior contextual understanding, which enable it to outperform the constrained or context-blind methodologies of existing tools.

\section{Our Approach}
\label{approach}


\begin{table*}[t]
	\centering
	\caption{Effectiveness of NLP2REST and RESTGPT.}
    \vspace*{-10pt}
	\label{tab:nlp2rest}
	\resizebox{\linewidth}{!}{
		\begin{tabular}{l|c|c|c|c|c|c|c|c|c|c|c|c|c|c|c|c|c|c|c}
			\toprule
			& \textbf{No. of Rules in} & \multicolumn{6}{c}{\textbf{NLP2REST Without Validation Process}} & \multicolumn{6}{|c}{\textbf{NLP2REST With Validation Process}} & \multicolumn{6}{|c}{\textbf{RESTGPT}} \\
			\textbf{REST Service} & \textbf{Ground Truth} & \textbf{TP} & \textbf{FP} & \textbf{FN} & \textbf{Precision} & \textbf{Recall} & \textbf{F\textsubscript{1}} & \textbf{TP} & \textbf{FP} & \textbf{FN} & \textbf{Precision} & \textbf{Recall} & \textbf{F\textsubscript{1}} & \textbf{TP} & \textbf{FP} & \textbf{FN} & \textbf{Precision} & \textbf{Recall} & \textbf{F\textsubscript{1}} \\ \midrule
            FDIC & 45 & 42 & 36 & 3 & 54\% & 93\% & 68\% & 42 & 25 & 3 & 63\% & 93\% & 75\% & 44 & 0 & 1 & 100\% & 98\% & 99\%\\
            Genome Nexus & 81 & 79 & 3 & 2 & 96\% & 98\% & 97\% & 79 & 3 & 2 & 96\% & 98\% & 97\% & 75 & 0 & 6 & 100\% & 93\% & 96\%\\
            LanguageTool & 20 & 20 & 12 & 0 & 63\% & 100\% & 77\% & 18 & 2 & 2 & 90\% & 90\% & 90\% & 18 & 0 & 3 & 100\% & 86\% & 92\%\\
            OCVN & 17 & 15 & 2 & 2 & 88\% & 88\% & 88\% & 13 & 1 & 4 & 93\% & 76\% & 84\% & 15 & 2 & 1 & 88\% & 94\% & 91\%\\
            OhSome & 14 & 13 & 66 & 1 & 16\% & 93\% & 28\% & 12 & 11 & 2 & 52\% & 80\% & 63\% & 12 & 3 & 2 & 80\% & 86\% & 83\%\\
            OMDb & 2 & 2 & 0 & 0 & 100\% & 100\% & 100\% & 2 & 0 & 0 & 100\% & 100\% & 100\% & 2 & 0 & 0 & 100\% & 100\% & 100\%\\
            REST Countries & 32 & 28 & 1 & 4 & 97\% & 88\% & 92\% & 28 & 0 & 4 & 100\% & 88\% & 93\% & 30 & 0 & 2 & 100\% & 94\% & 97\%\\
            Spotify & 88 & 83 & 68 & 5 & 55\% & 94\% & 69\% & 82 & 28 & 6 & 75\% & 93\% & 83\% & 86 & 2 & 4 & 98\% & 96\% & 97\%\\
            YouTube & 34 & 30 & 126 & 4 & 19\% & 88\% & 32\% & 28 & 9 & 6 & 76\% & 82\% & 79\% & 24 & 2 & 8 & 92\% & 75\% & 83\%\\
            \midrule
			\textbf{Total} & \textbf{333} & \textbf{312} & \textbf{314} & \textbf{21} & \textbf{50\%} & \textbf{94\%} & \textbf{65\%} & \textbf{304} & \textbf{79} & \textbf{29} & \textbf{79\%} & \textbf{91\%} & \textbf{85\%} & \textbf{306} & \textbf{9} & \textbf{27} & \textbf{97\%} & \textbf{92\%} & \textbf{94\%}\\ \bottomrule
		\end{tabular}
	}
\vspace*{-8pt}
\end{table*}

\begin{table}[ht!]
\centering
\caption{Accuracy of ARTE and RESTGPT.}
\vspace*{-10pt}
\resizebox{.5\columnwidth}{!}{
\begin{tabular}{lcc}
\toprule
\textbf{Service Name} & \textbf{ARTE} & \textbf{RESTGPT} \\
\midrule
FDIC & 25.35\% & 77.46\% \\
Genome Nexus & 9.21\% & 38.16\% \\
Language-Tool & 0\% & 82.98\% \\
OCVN & 33.73\% & 39.76\% \\
OhSome & 4.88\% & 87.80\% \\
OMDb & 36.00\% & 96.00\% \\
REST-Countries & 29.66\% & 92.41\% \\
Spotify & 14.79\% & 76.06\% \\
Youtube & 0\% & 65.33\% \\
\midrule
\textbf{Average} & 16.93\% & 72.68\% \\
\bottomrule
\end{tabular}
}
\label{tab:accuracy_comparison}
\end{table}

\subsection{Overview}
Figure~\ref{fig:overview} illustrates the RESTGPT workflow, which starts by parsing the input OpenAPI specification. During this phase, both machine-readable and human-readable sections of each parameter are identified. The human-readable sections provide insight into four constraint types: operational constraints, parameter constraints, parameter type and format, and parameter examples~\cite{kim2023enhancing}. 

The Rule Generator, using a set of crafted prompts, extracts these four rules. We selected GPT-3.5 Turbo as the LLM for this work, given its accuracy and efficiency, as highlighted in a recent report by OpenAI~\cite{openai2023gpt4}. 
The inclusion of few-shot learning further refines the model's output. By providing the LLM with concise, contextually-rich instructions and examples, the few-shot prompts ensure the generated outputs are both relevant and precise~\cite{brown2020language,liu2023pre}. Finally, RESTGPT combines the generated rules with the original specification to produce an enhanced specification.

\subsection{Rule Generator}


To best instruct the model on rule interpretation and output formatting, our prompts are designed around four core components: guidelines, cases, grammar highlights, and output configurations.  

\vspace*{-5pt}
\begin{tcolorbox}[title=Guidelines, colback=white, colframe=black!75, colbacktitle=black!85, coltitle=white]
1. Identify the parameter using its name and description.\\
2. Extract logical constraints from the parameter description, adhering strictly to the provided format.\\
3. Interpret the description in the least constraining way.
\end{tcolorbox}
\vspace*{-5pt}

The provided guidelines serve as the foundational instructions for the model, framing its perspective and clarifying its primary objectives. Using the guidelines as a basis, RESTGPT can then proceed with more specific prompting.

\vspace*{-5pt}
\begin{tcolorbox}[title=Cases, colback=white, colframe=black!75, colbacktitle=black!85, coltitle=white]
\textbf{Case 1:} If the description is non-definitive about parameter requirements: Output "None".\\
...\\
\textbf{Case 10:} For complex relationships between parameters: Combine rules from the grammar.
\end{tcolorbox}
\vspace*{-5pt}

The implementation of cases in model prompting plays a pivotal role in directing the model's behaviour, ensuring that it adheres to precise criteria as depicted in the example. Drawing inspiration from Chain-of-Thought prompting~\cite{wei2022chain}, we decompose rule extraction into specific, manageable pieces to mitigate ambiguity and, consequently, improve the model's processing abilities.

\vspace*{-5pt}
\begin{tcolorbox}[title=Grammar Highlights, colback=white, colframe=black!75, colbacktitle=black!85, coltitle=white]
\textbf{Relational Operators:} '$<$', '$>$', '$<=$', '$>=$', '$==$', '$!=$'\\
\textbf{Arithmetic Operators:} '$+$', '$-$', '$*$', '$/$'\\
\textbf{Dependency Operators:} 'AllOrNone', 'ZeroOrOne', ...
\end{tcolorbox}
\vspace*{-5pt}

The Grammar Highlights emphasize key operators and vocabulary that the model should recognize and employ during rule extraction. By providing the model with a fundamental context-specific language, RESTGPT identifies rules within text.

\vspace*{-5pt}
\begin{tcolorbox}[title=Output Configurations, colback=white, colframe=black!75, colbacktitle=black!85, coltitle=white]
\textbf{Example Parameter Constraint:} min [minimum], max [maximum], default [default]\\
\textbf{Example Parameter Format:} type [type], items [item type], format [format], collectionFormat [collectionFormat]
\end{tcolorbox}
\vspace*{-5pt}

After guiding the model through the rule-extraction process via specific prompting, we lastly define output formatting to compile the model's findings into a simple structure for subsequent processing.

Additionally, the Rule Generator also oversees the value-generation process, which is executed during the extraction of parameter example rules. Our artifact~\cite{artifact, zenodo} provides details of all the prompts and their corresponding results.

\subsection{Specification Enhancement}

The primary objective of RESTGPT is to improve the effectiveness of REST API testing tools. We accomplish this by producing enhanced OpenAPI specifications, augmented with rules derived from the human-readable natural-language descriptions in conjunction with the machine-readable OpenAPI keywords~\cite{oasdatamodel}.

As illustrated in Figure~\ref{fig:overview}, the \textit{Specification Parsing} stage extracts the machine-readable and human-readable components from the API specification. After rules from the natural language inputs have been identified by the \textit{Rule Generator}, the \textit{Specification Building} phase begins. During this phase, the outputs from the model are processed and combined with the machine-readable components, ensuring that there is no conflict between restrictions. For example, the resulting specification must have the \texttt{\small style} attribute only if the data type is \texttt{\small array} or \texttt{\small object}. The final result is an enriched API specification that contains constraints, examples, and rules extracted from the human-readable descriptions.

\section{Preliminary Results}
\label{evaluation}


\subsection{Evaluation Methodology}

We collected nine RESTful services from the NLP2REST study. The motivation behind this selection is the availability of a ground truth of extracted rules in the NLP2REST work~\cite{kim2023enhancing}. Having this data, we could easily compare our work with NLP2REST.

To establish a comprehensive benchmark, we incorporated a comparison with ARTE as well. Our approach was guided by the ARTE paper, from which we extracted the necessary metrics for comparison. Adhering to ARTE's categorization of input values as Syntactically Valid and Semantically Valid~\cite{alonso2022arte}, two of the authors meticulously verified the input values generated by RESTGPT and ARTE. Notably, we emulated ARTE's approach in scenarios where more than ten values were generated by randomly selecting ten from the pool for analysis.


\balance
\subsection{Results and Discussion}

Table~\ref{tab:nlp2rest} presents a comparison of the rule-extraction capabilities of NLP2REST and RESTGPT. RESTGPT excels in precision, recall, and the F\textsubscript{1} score across a majority of the REST services. NLP2REST, while effective, hinges on a validation process that involves evaluating server responses to filter out unsuccessful rules. This methodology demands engineering effort, and its efficacy is constrained by the validator's performance.

In contrast, RESTGPT eliminates the need for such validation entirely with its high precision. Impressively, RESTGPT's precision of 97\% surpasses even the precision of NLP2REST post-validation, which stands at 79\%. This emphasizes that RESTGPT is able to deliver superior results without a validation stage. This result shows an LLM's superior ability in nuanced rule detection, unlike conventional NLP techniques that rely heavily on specific keywords.

Furthermore, Table \ref{tab:accuracy_comparison} presents data on accuracy of ARTE and RESTGPT. The data paint a clear picture: RESTGPT consistently achieves higher accuracy than ARTE across all services. This can be attributed to the context-awareness capabilities of LLMs, as discussed in Section~\ref{background}. For example, in language-tool service, we found that, for the \texttt{\small language} parameter, ARTE generates values such as ``Arabic'', ``Chinese'', ``English'', and ``Spanish''. However, RESTGPT understands the context of the language parameter, and generates language code such as ``en-US'' and ``de-DE''.

\section{Future Plans}

\label{future}

Given our encouraging results on LLM-based rule extraction, we next outline several research directions that we plan to pursue in leveraging LLMs for improving REST API testing more broadly. 

\textbf{Model Improvement.} There are two ways in which we plan to create improved models for supporting REST API testing. First, we will perform task-specific fine-tuning of LLMs using data from APIs-guru~\cite{apis_guru} and RapidAPI~\cite{rapidapi}, which contain thousands of real-world API specifications.
We will fine-tune RESTGPT with these datasets, which should enhance the model's capability to comprehend diverse API contexts and nuances. We believe that this dataset-driven refinement will help RESTGPT understand a broader spectrum of specifications and generate even more precise testing suggestions.
Second, we will focus on creating lightweight models for supporting REST API testing, such that the models do not require expensive computational resources and can be deployed on commodity CPUs. To this end, we will explore approaches for trimming the model, focusing on retaining the essential neurons and layers crucial for our task.



\textbf{Improving fault detection.} RESTGPT is currently restricted to detecting faults that manifest as 500 server response codes. By leveraging LLMs, we intend to expand the types of bugs that can be detected, such as bugs related to CRUD semantic errors or discrepancies in producer-consumer relationships. By enhancing RESTGPT's fault-finding ability in this way, we aim to make automated REST API testing more effective and useful in practice.


\textbf{LLM-based Testing Approach.} We aim to develop a REST API testing tool that leverages server messages. Although server messages often contain valuable information, current testing tools fail to leverage this information~\cite{kim2023enhancing}. For instance, if a server hint suggests crafting a specific valid request, RESTGPT, with its semantic understanding, could autonomously generate relevant tests. This would not only enhance the testing process but also ensure that potential loopholes that the server messages may indicate would not be overlooked.

\section*{Acknowledgments}
  This work was partially supported by 
  NSF, under grant CCF-0725202,
  DOE, under contract DE-FOA-0002460,
  and gifts from Facebook, Google, IBM Research, and Microsoft Research.

\bibliographystyle{ACM-Reference-Format}
\bibliography{bib}

\end{document}